\documentstyle[aps]{revtex}%

\begin{document}%
\draft%

\title{Topological defect system in $O(n)$ symmetric time-dependent Ginzburg-Landau model}%
\author{Ying Jiang\thanks{E-mail: yjiang@itp.ac.cn}}%
\address{CCAST (World Laboratory), Box 8730, Beijing 100080, P. R. China}%
\address{Institute of Theoretical Physics, Chinese Academy of Sciences, P. O. Box 2735,
Beijing 100080, P. R. China\thanks{mailing address}}%
\maketitle%

\begin{abstract}
We present a new generalized topological current in terms of the
order parameter field $\vec \phi$ to describe the topological defect system
in $O(n)$ symmetric time-dependent Ginzburg-Landau model. With the aid of the $%
\phi$-mapping method, the structure of the topological defects and
the topological quantization of their topological charges in TDGL
model are obtained under the condition that the Jacobian $%
J(\frac \phi v)\neq 0$. We show that the topological defects are
generated from the zero points of the order parameter field $\vec
\phi$, and the topological charges of these topological defects
are topological quantized in terms of the Hopf indices and Brouwer
degrees of $\phi$-mapping under the condition. When $J(\frac \phi
v)=0$, it is shown that there exist the crucial case of branch
process. Based on the implicit function theorem and the Taylor
expansion, we detail the bifurcation of generalized topological
current and find different directions of the bifurcation. The
topological defects in TDGL model are found splitting or merging
at the degenerate point of field function $\vec \phi $ but the
total charge of the topological defects is still unchanged.
\end{abstract}


\section{Introduction}

The importance of the role of time-dependent Ginzburg-Landau
(TDGL) model in understanding a variety of problems in physics
is clear\cite{2}. The study of the growth kinetics of
systems subjected to rapid temperature quenches\cite{26,68,80}
has recently been extended to include systems with more complex
order parameter symmetries\cite{100,105}. In particular there
has been progress on the study of the $n$-dimensional vector
model with nonconserved Langevin dynamics. An interface
description and numerical simulations of a TDGL equation were
used to investigate the intra-surface kinetics of phase ordering
on toroidal and corrugated surfaces\cite{93}. Two-dimensional
$XY$ models with resistively-shunted junction (RSJ) dynamics and
TDGL dynamics were simulated and it was verified that the vortex
responses can be well described by the Minnhagen phenomenology
for both types of dynamics\cite{70}.

In recent years, much work has been done on the topological
defects in the TDGL model\cite{2,1,3,4}. In certain
cosmological\cite{18} and phase ordering\cite{11} problems key
questions involve an understanding of the evolution and
correlation among defects like vortices, monopoles, domain walls,
etc. In studying such objects in field theory questions arise as
to how one can define quantities like the density of topological
defects and an associated velocity field. Liu and
Mazenko\cite{2,1} have discussed the problem, but unfortunately,
for the lack of a powerful method, the topological properties of
these systems are not very clear, some important topological
information has been lost, also the unified theory of describing
the topological properties of all defect system in TDGL model is
not established yet.

In our previous work\cite{jiangpla}, we discussed how one could
use the $\phi$-mapping topological current theory to study the
topological structure of the point defects in phase-ordering
systems. We deduced a topological current of point-defect system
in terms of the order parameter field in the context of a
$d$-dimensional $O(n)$ symmetric TDGL model, where $d$ is the
spatial dimensionality. Using the expression of the topological
current, for point defects ($n=d$), the point-defect velocity
field was identified, the topological structure and the
topological charges of the point-defect system were studied, also
the branch process of the point defects was discussed
systematically. This analysis will be extended here to the case of
arbitrary dimensional defect system in TDGL model where $n=d-k$,
i.e. to study the case for all $n\leq d$.

In this paper, in the light of $\phi $--mapping topological current theory%
\cite{duan3}, a useful method which plays a important role in
studying the topological invariants\cite{duan4,duan8} and the
topological structures of physical systems\cite{duan6,duan9}, we
will investigate the topological quantization and the branch
process of arbitrary dimensional topological defects in $O(n)$
symmetric TDGL model. We will show that the topological defects
are generated from the zero points of the order parameter field
$\vec \phi$, and their topological
charges are quantized in terms of the Hopf indices and Brouwer degrees of $%
\phi$-mapping under the condition that the zero points of field
$\vec \phi$ are regular points. While at the critical points of
the order parameter field $\vec{\phi}$, i.e. the limit points and
bifurcation points, there exist branch processes, the topological
current of defect bifurcates and the topological defects split or
merge at such point, this means that the topological defects
system is unstable at these points.

This paper is organized as follows. In section 2, we investigate
the topological structure of the defect system in TDGL model and
give the expression of the defect velocity field. In section 3, we
will point out that the topological charges of these defects are
the Winding numbers which are
determined by the Hopf indices and the Brouwer degrees of the $\phi $%
--mapping. In section 4, we study the branch process of the defect
topological current at the limit points, bifurcation points and
higher degenerated points systematically by virtue of the $\phi
$--mapping theory and the implicit function theorem.

\section{Topological structure of the defect
system in TDGl model}

We consider a TDGL model for an $n$-component order parameter
$\vec \phi=(\phi ^1 (\vec r, t),...,\phi ^n (\vec r, t))$ in
$d$-spatial dimensions ($d-n=k$) governed by the Langevin equation
\begin{equation}\label{1}
  \frac{\partial\vec{\phi}}{\partial
  t}=\vec{K}\equiv-\Gamma\frac{\delta F}{\delta\vec{\phi}}+\vec{\eta}
\end{equation}
where $\Gamma$ is a kinetic coefficient and $\vec \eta$ is a
thermal noise which is related to $\Gamma$ by
fluctuation-dissipation theorem. $F$ is a Ginzburg-Landau
effective free energy assumed to be of the form
\begin{equation}\label{2}
  F=\int d^dr[\frac12c(\nabla\vec{\phi})^2+V(\vec \phi)]
\end{equation}
where $c>0$ and the potential $V$ is assumed to be of the
degenerate double-well form. This model is to be supplemented by
random, uncorrelated, initial conditions. We assume that there is
a rapid temperature quench from a high temperature to zero
temperature where the noise $\vec \eta$ in (\ref{1}) can be set to
zero. In the scalar case ($n=1$) such system order through the
growth of domains separated by sharp walls. As time evolves these
domains coarsen and order grows to progressively longer length
scales. In the case of systems with continuous symmetry ($n>1$)
the disordering elements\cite{a,b} will depend on $n$ and spatial
dimensionality $d$. Thus, for example, for $n=d$ one has point
defects (vortices and monopoles),for $n=d-1$ one has vortex line
or string-like defects while for $n=d-k$ one has $k$-dimensional
topological defect objects. For $n>d$ there are no stable singular
topological objects.

As is well known, the $n$-component order parameter field $\vec
\phi(\vec r, t)$ determines the defect properties of the system,
and it can be looked upon as a smooth mapping between the
$(d+1)$-dimensional space-time $X$ and the $n$-dimensional
Euclidean space $R^n$ as $\phi : X \rightarrow R^n$. By analogy
with the discussion in our previous
work\cite{jiangpla,duan3,duan9,jiang1}, from this $\phi$-mapping,
one can deduce a topological tensor current as
\[
  j^{\mu _0 \mu _1\cdot \cdot \cdot \mu _k}=\frac 1{A(S^{n-1})(n-1)!}%
\epsilon ^{\mu _0 \mu _1\cdot \cdot \cdot \mu _k\mu _{k+1}\cdot
\cdot \cdot \mu _d}\epsilon _{a_1\cdot \cdot \cdot a_n}\partial
_{\mu _{k+1}}n^{a_1}\partial _{\mu _{k+2}}n^{a_2}\cdot \cdot \cdot
\partial _{\mu _d}n^{a_n},
\]
\begin{equation}\label{3}
\mu _0,...,\mu _d=0,1,...,d, \;\;\;a_1,...,a_n=1,...,n
\end{equation}
to describe the $k$-dimensional topological defects in TDGL model,
and the density tensor of the topological defect system is defined
as $\rho ^{\mu _1\cdot \cdot \cdot \mu _k}=j^{0\mu _1\cdot \cdot
\cdot \mu _k}$. In this expression, $\partial_\mu$ stands for
$\partial/
\partial x^\mu$, $A(S^{n-1})=2\pi ^{n/2}/\Gamma (n/2)$ is the area
of $(n-1)$--dimensional unit sphere $S^{n-1}$ and $n^a (x)$ is the
direction field of the $n$--component order parameter field $\vec
\phi$
\begin{equation}
n^a(x)=\frac{\phi ^a(x)}{||\phi (x)||},\;\;\;\;\;||\phi
(x)||=\sqrt{\phi ^a(x)\phi ^a(x)}  \label{4}
\end{equation}
with $n^a(x)n^a(x)=1$. It is obviously that $n^a(x)$ is a section
of the sphere bundle $S(X)$\cite{duan3} and it can be looked upon
as a map of $X$ onto a $(n-1)$--dimensional unit sphere $S^{n-1}$
in order parameter space. Clearly, the zero points of the order
parameter field $\vec{\phi}(x)$ are just the singular points of
the unit vector $n^a(x)$. It is easy to see that $j^{\mu _0\cdot
\cdot \cdot \mu _k}$ are completely antisymmetric, and from the
formulas above, we conclude that there exists a conservative
equation of the topological tensor current in (\ref{3})
\[
\partial _{\mu _i}j^{\mu _0\cdot \cdot \cdot \mu
_k}=0,\;\;\;i=0,...,k,
\]
and from which we have
\[
\partial _t \rho ^{\mu_1 \cdot \cdot \cdot \mu_k}+\nabla_\mu
j^{\mu \mu_1 \cdot \cdot \cdot \mu_k}=0
\]
which is just the continuity equation satisfied by $\rho ^{\mu_1
\cdot \cdot \cdot \mu_k}$.

In the following, we will investigate the intrinsic structure of
the generalized topological current $j^{\mu _0\mu _1\cdot \cdot
\cdot \mu _k}$ by making use of the $\phi $--mapping method. From
(\ref{4}), we have
\[
\partial _\mu n^a=\frac 1{||\phi ||}\partial _\mu \phi ^a+\phi ^a\partial
_\mu (\frac 1{||\phi ||}),\;\;\;\frac \partial {\partial \phi ^a}(\frac 1{%
||\phi ||})=-\frac{\phi ^a}{||\phi ||^3}
\]
which should be looked upon as generalized
functions\cite{gelfand1}. Due to these expressions the generalized
topological current (\ref{3}) can be rewritten as
\begin{eqnarray}
j^{\mu _0\mu _1\cdot \cdot \cdot \mu _k} &=&C_n\epsilon ^{\mu
_0\mu _1\cdot \cdot \cdot \mu _k\mu _{k+1}\cdot \cdot \cdot \mu
_d}\epsilon _{a_1\cdot \cdot \cdot a_n}  \nonumber \\
&\cdot& \partial _{\mu _{k+1}}\phi ^a\cdots \partial _{\mu _d}\phi ^{a_n}%
\frac \partial {\partial \phi ^a}\frac \partial {\partial \phi ^{a_1}}%
(G_n(||\phi ||)),
\end{eqnarray}
where $C_m$ is a constant
\[
C_n=\left\{
\begin{array}{cc}
-\frac 1{A(S^{n-1})(n-2)(n-1)!}, & \;\;\;\;n>2 \\ \frac 1{2\pi },
& \;\;\;\;n=2
\end{array}
,\right.
\]
and $G_n(||\phi ||)$ is a Green function
\[
G_n(||\phi ||)=\left\{
\begin{array}{ccc}
\frac 1{||\phi ||^{n-2}} & ,\;\;\; & n>2 \\ \ln ||\phi || &
,\;\;\; & n=2
\end{array}
.\right.
\]
Defining general Jacobians $J^{\mu _0\mu _1\cdot \cdot \cdot \mu
_k}(\frac \phi x)$ as following
\[
\epsilon ^{a_1\cdot \cdot \cdot a_n}J^{\mu _0\mu _1\cdot \cdot \cdot \mu _k}(\frac %
\phi x)=\epsilon ^{\mu _0\mu _1\cdot \cdot \cdot \mu _k\mu
_{k+1}\cdot \cdot \cdot \mu _d}\partial _{\mu _{k+1}}\phi
^{a_1}\partial _{\mu _{k+2}}\phi ^{a_2}\cdot \cdot \cdot \partial
_{\mu _d}\phi ^{a_n}
\]
and by making use of the $n$--dimensional Laplacian Green function
relation in $\phi$--space\cite{duan3}
\[
\Delta _\phi (G_n(||\phi ||))=-\frac{4\pi ^{n/2}}{\Gamma (\frac n2%
-1)}\delta (\vec{\phi})
\]
where $\Delta _\phi =(\frac{\partial ^2}{\partial \phi ^a\partial
\phi ^a})$ is the $n$--dimensional Laplacian operator in $\phi
$--space, we do obtain the $\delta $--function structure of the
defect topological current rigorously
\begin{equation}
j^{\mu _0\mu _1\cdot \cdot \cdot \mu _k}=\delta (\vec{\phi}%
)J^{\mu _0\mu _1\cdot \cdot \cdot \mu _k}(\frac \phi x).
\label{4.10}
\end{equation}
This expression involves the total defect information of the
system in TDGL model and it indicates that all the defects are
located at the zero points of the order parameter field $\vec \phi
(x)$. It must be pointed out that, comparing to similar
expressions in other papers, the results in (\ref{4.10}) is gotten
theoretically in a natural way. From this expression, the density
tensor of $j^{\mu _0 \mu _1\cdot \cdot \cdot \mu _k}$ is also
changed into a compact form
\[
\rho ^{\mu _1 \cdot \cdot \cdot \mu_k}=j^{0\mu _1 \cdot \cdot
\cdot \mu_k}=\delta (\vec{\phi})D^{\mu _1\cdot \cdot \cdot \mu
_k}(\frac \phi x),
\]
where $D^{\mu _1\cdot \cdot \cdot \mu _k}(\frac \phi x)=J^{0\mu
_1\cdot \cdot \cdot \mu _k}(\frac \phi x)$. We find that $j^{\mu
_0 \mu _1\cdot \cdot \cdot \mu _k}\neq 0$, $\rho^{\mu _1\cdot
\cdot \cdot \mu _k}\neq 0$
only when $\vec{\phi}%
=0$, which is just the singularity of $j^{\mu _0\mu _1\cdot \cdot
\cdot \mu _k}$. In detail, the Kernel of the $\phi $--mapping is
the singularities of the topological tensor current $j^{\mu _0\mu
_1\cdot \cdot \cdot \mu _k}$ in $X$, i.e. the inner structure of
the topological tensor current is labeled by the zeroes of
$\phi$-mapping. We think that this is the essential of the
topological tensor current theory and $\phi $--mapping is the key
to study this theory.

From the above discussions, we see that the kernel of
$\phi$--mapping plays an important role in the topological tensor
current theory, so we are focused on the zero points of $\vec
\phi$ and will search for the solutions of the equations $\phi^a
(x)=0$ $(a=1,...,n)$ by means of the implicit function theorem.
These points are topological singularities in the orientation of
the order parameter field $\vec \phi (x)$.

Suppose that the vector field $\vec{\phi}(x)$ possesses $l$
zeroes, according to the implicit function theorem\cite
{goursat1}, when the zeroes are regular points of $\phi $--mapping
at which the rank of the Jacobian matrix $[\partial _\mu \phi ^a]$
is $n$, the solutions of $\vec{\phi}=0$ can be expressed
parameterizedly by
\begin{equation}
x^\mu =z_i^\mu (t, u^1,\cdot \cdot \cdot ,u^k),\;\;\;\;i=1,...,l,
\label{4.5}
\end{equation}
where the subscript $i$ represents the $i$--th solution and the parameters $%
u^I$ ($I=1,...,k$), combining with the time parameter $t$, span a
$(k+1)$--dimensional subspace which is called the $i$--th singular
subspace $N_i$ in the space-time $X$ corresponding to the $\phi $%
--mapping. These singular subspaces $N_i$ are just the world
volumes of the topological defects, the parameters $u^I$ play the
role of the spatial parameters of the topological defects. For
each singular subspace $N_i$, we can define a normal subspace
$M_i$ in $X$ which is spanned by the parameters $v^A$ ($A=1,\cdot
\cdot \cdot , n$), and the intersection point of $M_i$ and $N_i$
is denoted by $p_i$ which can be expressed parameterizedly by
$v^A=p_i^A$. In fact, in the words of differential topology, $M_i$
is transversal to $N_i$ at the point $p_i$. By virtue of the
implicit function theorem at the regular point $p_i$, it should be
held true that the Jacobian matrices $J(\frac \phi v)$ satisfies
$J(\frac \phi v)=\frac{D(\phi ^1,\cdot \cdot \cdot ,\phi
^n)}{D(v^1,\cdot \cdot \cdot ,v^n)}\neq 0$.

On the other hand, putting the solutions (\ref{4.5}) into $\vec
\phi (x)$, we have $\vec \phi (z_i)\equiv 0$, from this
expression, and since we expect the instantaneous velocity to be
orthogonal to the local orientation of the topological defects,
we can define the velocity via
\begin{equation}
J ^{\mu \mu_1 \cdot \cdot \cdot \mu _k}=v ^{[ \mu} D ^{\mu _1
\cdot \cdot \cdot \mu _k ]},
\end{equation}
that is, the topological current and the density tensor should
satisfy
\begin{equation}
j ^{\mu \mu_1 \cdot \cdot \cdot \mu_k}=v ^{[ \mu} \rho ^{\mu _1
\cdot \cdot \cdot \mu _k]}. \label{ff}
\end{equation}
This expression for the velocity can be used to find the defect
velocity distribution in the case of phase-ordering kinetics for a
non-conserved order parameter\cite{2,1,jiangpla}. It is very
useful because it avoid the problem of having to specify the
positions of the topological defects explicitly. The positions are
implicitly determined by the zeros of the order parameter field.
The general expression with $J ^{\mu \mu _1 \cdot \cdot \cdot \mu
_k} (\phi /x)$ should be useful in looking at the motion of the
topological defects in TDGL model in the presence of external
fields beyond a growth kinetics context\cite{2,1}.

If we restrict the discussion to the simplest case of point
defects ($n=d$), the corresponding topological current is
reduced to $j ^\mu =\delta (\phi)J ^\mu (\phi /x)$, and
Eq.(\ref{ff}) can be put into a conventional form $j ^\mu =v
^\mu \rho$ with $v ^\mu$ taking the expression of $v ^\mu =J
^\mu (\phi /x)/D(\phi /x)$\cite{jiangpla}. Also, the string in
TDGl model, which was discussed by Mazenko\cite{mazenko3}, is a
special case for $n=d-1$ in our theory.

\section{Topological quantization of the defect charges in TDGL
model}

In the following, we will investigate the topological charges of
the topological defects and their quantization. Let $\Sigma _i$ be
a neighborhood of $p_i$ on $M_i$ with boundary $\partial \Sigma
_i$ satisfying $p_i\notin \partial \Sigma_i $, $\Sigma _i\cap
\Sigma _j=\emptyset $. Then the generalized winding number $W_i$
of $n^a (x)$ at $p_i$\cite{dubrosin1} can be defined by the Gauss
map $n:\partial \Sigma _i\rightarrow S^{n-1}$
\begin{equation}
W_i=\frac 1{A(S^{n-1})(n-1)!}\int_{\partial \Sigma
_i}n^{*}(\epsilon _{a_1\cdot \cdot \cdot a_n}n^{a_1}dn^{a_2}\wedge
\cdot \cdot \cdot \wedge dn^{a_n})  \label{4.6}
\end{equation}
where $n^{*}$ denotes the pull back of map $n$. The generalized
winding numbers is a topological invariant and is also called the
degree of Gauss map\cite{milnor1}. It means that, when the point
$v^A$ covers $\partial \Sigma_i $ once, the unit vector $n^a$ will
cover a region $n[\partial \Sigma _i]$ whose area is $W_i$ times
of $A(S^{n-1})$, i.e. the unit vector $n^a$ will cover the unit
sphere $S^{n-1}$ for $W_i$ times. Using the Stokes' theorem in
exterior differential form and duplicating the above process, we
get the compact form of $W_i$
\begin{equation}
W_i=\int_{\Sigma _i}\delta (\vec{\phi})J(\frac \phi v)d^nv.
\label{4.W}
\end{equation}
By analogy with the procedure of deducing $\delta (f(x))$, since
\begin{equation}
\delta (\vec{\phi})=\left\{
\begin{array}{cc}
+\infty , & for\;\vec{\phi}(x)=0 \\ 0, & for\;\vec{\phi}(x)\neq 0
\end{array}
\right. =\left\{
\begin{array}{cc}
+\infty , & for\;x\in N_i \\ 0, & for\;x\notin N_i
\end{array}
\right. ,
\end{equation}
we can expand the $\delta $--function $\delta (\vec{\phi})$ as
\begin{equation}
\delta (\vec{\phi})=\sum_{i=1}^lc_i\delta (N_i),  \label{4.delta}
\end{equation}
where the coefficients $c_i$ must be positive, i.e. $c_i=\mid
c_i\mid $. $\delta (N_i)$ is the $\delta $--function in space-time
$X$ on a submanifold $N_i$\cite{gelfand1}
\begin{equation}
\delta (N_i)=\int_{N_i}\delta ^n(\vec{x}-\vec{z}%
_i(t,u^1,\cdot \cdot \cdot, u^k))dtd^ku.  \label{4.m}
\end{equation}
Substituting (\ref{4.delta}) into (\ref{4.W}), and calculating the
integral, we get the expression of $c_i$
\begin{equation}
c_i=\frac{\beta _i}{\mid J(\frac \phi v)_{p_i}\mid }=\frac{\beta
_i\eta _i}{J(\frac \phi v)_{p_i}},
\end{equation}
where $\beta _i=|W_i|$ is a positive integer called the Hopf
index\cite {milnor1} of $\phi $-mapping on $M_i,$ it means that
when the point $v$
covers the neighborhood of the zero point $p_i$ once, the function $\vec{\phi%
}$ covers the corresponding region in $\vec{\phi}$-space $\beta
_i$ times, and $\eta _i=signJ(\frac \phi v)_{p_i}=\pm 1$ is the
Brouwer degree of $\phi $-mapping\cite{milnor1}. Substituting this
expression of $c_i$ and (\ref {4.delta}) into (\ref{4.10}), we
gain the total expansion of the topological current
\[
j^{\mu _0\mu _1\cdot \cdot \cdot \mu _k}=\sum_{i=1}^l\frac{%
\beta _i\eta _i}{J(\frac \phi v)|_{p_i}}\delta (N_i)J^{\mu _0\mu
_1\cdot \cdot \cdot \mu _k}(\frac \phi x).
\]
or in terms of parameters $y^{A^{^{\prime }}}=(t, v^1,\cdot \cdot
\cdot
,v^n,u^1,\cdot \cdot \cdot ,u^k)$%
\begin{equation}
j^{A_0^{^{\prime }} A_1^{^{\prime }}\cdot \cdot \cdot A_k^{^{\prime }}}=%
\sum_{i=1}^l\frac{\beta _i\eta _i}{J(\frac \phi v)|_{p_i}}\delta
(N_i)J^{A_0^{^{\prime }} A_1^{^{\prime }}\cdot \cdot \cdot
A_k^{^{\prime }}}(\frac \phi y).
\end{equation}
From the above equation, we conclude that the inner structure of
$j^{\mu _0 \mu _1\cdot \cdot \cdot \mu _k}$ or $j^{A_0^{^{\prime
}} A_1^{^{\prime }}\cdot \cdot \cdot
A_k^{^{\prime }}}$ is labeled by the total expansion of $\delta (\vec{\phi}%
) $, and it just represents $l$ $(k)$-dimensional topological
defects with topological charges $g_i=\beta _i \eta _i$ moving in
the $(d+1)$--dimensional
space-time $X$. The $(k+1)$-dimensional singular subspaces $%
N_i\,\,(i=1,\cdot \cdot \cdot l)$ are their world sheets in the
space-time. Mazenko\cite{2,1} and Halperin\cite{halperin} also got
similar results for the case of point-like defects and line
defects, but unfortunately, they did not consider the case $\beta
_i\neq 1$. In fact, what they lost sight of is just the most
important topological information for the charge of topological
defects. In detail, the Hopf indices $\beta _i$ characterize the
absolute values of the topological charges of these defects and
the Brouwer degrees $\eta _i=+1$ correspond to defects while $\eta
_i=-1$ to antidefects. Furthermore, they did not discuss what will
happen when $\eta _i$ is indefinite, which we will study in detail
in section 4.

\section{The branch process of the topological current}

With the discussion mentioned above, we know that the results in
the above section are obtained straightly from the topological
view point under the condition $J(\phi/v)|_{p_i}\neq 0$, i.e. at
the regular points of the order parameter field $\vec \phi$. When
the condition fails, i.e. the Brouwer degree $\eta_i$ are
indefinite, there should exist some kind of branch processes in
the topological current of the topological defect system in TDGL
model. In what follows, we will study the case when
$J(\phi/v)|_{p_i}=0$. It often happens when the zero points of
field $\vec \phi$ include some branch points, which lead to the
bifurcation of the topological current.

In this section, we will discuss the branch processes of these
topological defects. In order to simplify our study,
let the spatial parameters $%
u^I$ be fixed, i.e. to choose a fixed point on
the topological defect. In this case, the Jacobian matrices $%
J^{A_0^{^{\prime }} A_1^{^{\prime }}\cdot \cdot \cdot
A_k^{^{\prime }}}(\frac \phi y)$ are reduced to
\[
J^{AI_1\cdot \cdot \cdot I_{k}}(\frac \phi y)\equiv J^A(\frac \phi y%
),\;\;\;\;J^{ABI_1\cdot \cdot \cdot I_{k-1}}(\frac \phi y)=0,\;\;\;%
\;J^{(n+1)\cdot \cdot \cdot d}(\frac \phi y)=J(\frac \phi v),
\]
\begin{equation}
A,B=0,1,...,n,\;\;\;\;I_j=n+1,...,d,
\end{equation}
for $y^A=v^A\;(A\leq n),\;y^{0}=t,\;y^{n+I}=u ^I\;(I\geq 1)$. The
branch points are determined by the $n+1$ equations
\begin{equation}
\phi ^a(t,v^1,\cdots ,v^n,\vec{u})=0,\;\;\;a=1,...,n
\label{4.phia}
\end{equation}
and
\begin{equation}
\phi ^{n+1}(t,v^1,\cdots ,v^n,\vec{u})\equiv J(\frac \phi v)=0
\label{4.zero}
\end{equation}
for the fixed $\vec{u}$. and they are denoted as $(t^{*},p_i)$. In $%
\phi $--mapping theory usually there are two kinds of branch
points, namely the limit points and bifurcation
points\cite{kubicek1}, satisfying
\begin{equation}
J^1(\frac \phi y)|_{(t^{*},p_i)}\neq 0  \label{4.nonzero1}
\end{equation}
and
\begin{equation}
J^1(\frac \phi y)|_{(t^{*},p_i)}=0,  \label{4.zero1}
\end{equation}
respectively. In the following, we assume that the branch points $%
(t^{*},p_i) $ of $\phi $--mapping have been found.

\subsection{The branch process at the limit point}

We first discuss the branch process at the limit point satisfying
the condition (\ref{4.nonzero1}). In order to use the theorem of
implicit function to study the branch process
of topological defects at the limit point, we use the Jacobian $J^1(\frac %
\phi y)$ instead of $J(\frac \phi v)$ to discuss the problem. In
fact, this means that we have replaced the parameter $t$ by $v^1$.
Then, taking account of the condition(\ref{4.nonzero1}) and using
the implicit function theorem, we have an unique solution of the
equations (\ref
{4.phia}) in the neighborhood of the limit point $(t^{*},p_i)$%
\begin{equation}
t=t(v^1,\vec{u}),\;\;\;\;v^i=v^i(v^1,\vec{u}),\;\;\;\;i=2,3,...,n
\label{4.102}
\end{equation}
with $t^{*}=t(p_i^1,\vec{u})$. In order to show the behavior of
the
defects at the limit points, we will investigate the Taylor expansion of (%
\ref{4.102}) in the neighborhood of $(t^{*},p_i)$. In the present
case, from (\ref{4.nonzero1}) and (\ref{4.zero}), we get
\[
\frac{dv^1}{dt}|_{(t^{*},p_i)}=\frac{J^1(\frac \phi y)}{J(\frac \phi v)}%
|_{(t^{*},p_i)}=\infty ,
\]
i.e.
\[
\frac{dt}{dv^1}|_{(t^{*},p_i)}=0.
\]
Therefore, the Taylor expansion of (\ref{4.102}) at the point
$(t^{*},p_i)$ gives
\begin{equation}
t-t^{*}=\frac 12\frac{d^2t}{(dv^1)^2}|_{(t^{*},p_i)}(v^1-p_i^1)^2
\label{4.103}
\end{equation}
which is a parabola in the $v^1$---$t$ plane. From (\ref{4.103}),
we can
obtain the two solutions $v_{(1)}^1(t,\vec{u})$ and $v_{(2)}^1(t,\vec{%
u})$, which give the branch solutions of the system (\ref{4.phia})
at the limit point. If $\frac{d^2t}{(dv^1)^2}|_{(t^{*},p_i)}>0$,
we have the
branch solutions for $t>t^{*}$ (Fig 1(a)), otherwise, we have the branch solutions for $%
t<t^{*} $ (Fig 1(b)). The former is related to the creation of
defect and antidefect in pair at the limit points, and the latter
to the annihilation of the topological defects, since the
topological current of topological defects is identically
conserved, the topological quantum numbers of these two generated
topological defects must be opposite at the limit point, i.e.
$\beta _1\eta _1+\beta _2\eta _2=0$.

\subsection{The branch process at the bifurcation point}

In the following, let us consider the case (\ref{4.zero1}), in
which the
restrictions of the system (\ref{4.phia}) at the bifurcation point $%
(t^{*},p_i) $ are
\begin{equation}  \label{4.104}
J(\frac \phi v)|_{(t^{*},p_i)}=0,\;\;\;J^1(\frac \phi
y)|_{(t^{*},p_i)}=0.
\end{equation}
These two restrictive conditions will lead to an important fact
that the dependency relationship between $t$ and $v^1$ is not
unique in the neighborhood of the bifurcation point $(t^{*},p_i).$
In fact, we have
\begin{equation}  \label{4.105}
\frac{dv^1}{dt}|_{(t^{*},p_i)}=\frac{J^1(\frac \phi y)}{J(\frac \phi v)}%
|_{(t^{*},p_i)}
\end{equation}
which under the restraint (\ref{4.104}) directly shows that the
tangential direction of the integral curve of equation
(\ref{4.105}) is indefinite at the point $(t^{*},p_i)$. Hence,
(\ref{4.105}) does not satisfy the conditions of the existence and
uniqueness theorem of the solution of a differential equation.
This is why the very point $(t^{*},\vec z_i)$ is called the
bifurcation point of the system (\ref{4.phia}).

As we have mentioned above, at the bifurcation point $%
(t^{*},p_i)$, the rank of the Jacobian matrix $[\frac{\partial \phi }{%
\partial v}]$ is smaller than $n$. For the aim of searching for the different directions
of all branch curves at the bifurcation point, we firstly consider the rank of the Jacobian matrix $[\frac{\partial \phi }{%
\partial v}]$ is $n-1$. The case of a more smaller rank will be discussed in
next subsection. Let $J_1(\frac \phi v)=[\phi ^a _A] \; (a=1,\cdot
\cdot \cdot ,n-1; \; A=2, \cdot \cdot \cdot,n) $ be one of the
$(n-1)\times (n-1)$ submatrix of the Jacobian matrix
$[\frac{\partial \phi }{\partial v}]$ with $\det J_1(\frac \phi
v)\neq 0$
at the point $%
(t^{*},p_i)$ (otherwise, we have to rearrange the equations of (\ref{4.phia}%
)), where $\phi _A^a$ stands for $(\partial \phi ^a/\partial
v^A)$. By means of the implicit function theorem we obtain one and
only one functional relationship in the neighborhood of the
bifurcation point $(t^{*},p_i)$%
\begin{equation}  \label{4.107}
v^A=f^A(v^1,t,\vec u),\;\;\;\;\;A=2,3,...,n.
\end{equation}
We denote the partial derivatives as $f_1^A=\frac{\partial
v^A}{\partial v^1}$, $f_t^A=\frac{\partial v^A}{\partial t}$,
$f_{11}^A=\frac{\partial ^2v^A}{(\partial v^1)^2}$,
$f_{1t}^A=\frac{\partial ^2v^A}{\partial v^1\partial t}$,
$f_{tt}^A=\frac{\partial ^2v^A}{\partial t^2}$. From
(\ref{4.phia}) and (\ref{4.107}), we have for $a=1,...,n-1$
\[
\phi ^a=\phi ^a(v^1,f^2(v^1,t,\vec u ),...,f^n(v^1,t,\vec u ),t,%
\vec u )\equiv 0
\]
which gives
\begin{equation}  \label{4.108}
\sum\limits_{A=2}^n\frac{\partial \phi ^a}{\partial v^A}f_1^A=-\frac{%
\partial \phi ^a}{\partial v^1},\;\;\;a=1,...,n-1
\end{equation}
\begin{equation}  \label{4.109}
\sum\limits_{A=2}^n\frac{\partial \phi ^a}{\partial v^A}f_t^A=-\frac{%
\partial \phi ^a}{\partial t},\;\;\;a=1,...,n-1.
\end{equation}
By differentiating (\ref{4.108}) and (\ref{4.109}) with respect to
$v^1$ and $t$, and applying the Gaussian elimination method, we
can find the second order derivatives $f^A_{11}$, $f^A_{1t}$ and
$f^A_{tt}$. The above discussions do not matter to the last
component $\phi ^n(v^1,\cdot \cdot \cdot ,v^n,t,\vec u )$. In
order to find the different values of $dv^1/dt$ at the bifurcation
point, let us investigate the Taylor expansion of $\phi
^n(v^1,\cdot \cdot \cdot ,v^n,t,\vec u )$ in the neighborhood of
$(t^{*},p_i)$. Substituting (\ref{4.107}) into $\phi ^n(v^1,\cdot
\cdot \cdot
,v^n,t,\vec u )$, we get the function of two variables $v^1$ and $t$%
\begin{equation}  \label{4.113}
F(t,v^1,\vec u )=\phi ^m(v^1,f^2(v^1,t,\vec u
),...,f^m(v^1,t,\vec %
u ),t,\vec u )
\end{equation}
which according to (\ref{4.phia}) must vanish at the bifurcation
point
\begin{equation}  \label{4.114}
F(t^{*},p_i)=0.
\end{equation}
From (\ref{4.113}), we can calculate the first order partial derivatives of $%
F(t,v^1,\vec u )$ with respect to $v^1$ and $t$ respectively
at the
bifurcation point $(t^{*},p_i)$%
\begin{equation}  \label{4.115}
\frac{\partial F}{\partial v^1}=\phi _1^n+\sum\limits_{A=2}^n\phi
_A^nf_1^A,\;\;\;\frac{\partial F}{\partial t}=\phi
_t^n+\sum\limits_{A=2}^n\phi _A^nf_t^A.
\end{equation}
By making use of (\ref{4.108}) and (\ref{4.109}), with the
Cramer's rule, the first equation of (\ref{4.104}) is expressed as
\[
\frac{\partial F}{\partial v^1}\det J_1(\frac \phi
v)|_{(t^{*},p_i)}=0.
\]
Since $ \det J_1(\frac \phi v)|_{(t^{*},p_i)}\neq 0$, the above
equation leads to
\begin{equation}  \label{4.116}
\frac{\partial F}{\partial v^1}|_{(t^{*},p_i)}=0.
\end{equation}
With the same reasons, we can prove that
\begin{equation}  \label{4.117}
\frac{\partial F}{\partial t}|_{(t^{*},p_i)}=0.
\end{equation}
The second order partial derivatives of the function $F(t,v^1,\vec
u )$ are easily to find out from (\ref{4.115}) which at
$(t^{*},p_i)$ are denoted by
\begin{equation}  \label{4.118}
A=\frac{\partial ^2F}{(\partial v^1)^2}\mid _{(t^{*},p_i)},\quad \quad B=%
\frac{\partial ^2F}{\partial v^1\partial t}\mid _{(t^{*},p_i)},\
\quad \quad C=\frac{\partial ^2F}{\partial t^2}\mid
_{(t^{*},p_i)}.
\end{equation}
Then, by virtue of (\ref{4.114}), (\ref{4.116}), (\ref{4.117}) and
(\ref {4.118}), the Taylor expansion of $F(t,v^1,\vec u )$ in the
neighborhood of the bifurcation point $(t^{*},p_i)$ gives
\begin{equation}  \label{4.120}
A(v^1-p_i^1)^2+2B(v^1-p_i^1)(t-t^{*})+C(t-t^{*})^2=0.
\end{equation}
Dividing (\ref{4.120}) by $(v^1-p_i^1)^2$ or $(t-t^{*})^2$, and
taking the limit $t\rightarrow t^{*}$ as well as $v^1\rightarrow
p_i^1$ respectively, we get two equations
\begin{equation}  \label{4.121}
A(\frac{dv^1}{dt})^2+2B\frac{dv^1}{dt}+C=0.
\end{equation}
and
\begin{equation}  \label{4.122}
C(\frac{dt}{dv^1})^2+2B\frac{dt}{dv^1}+A=0.
\end{equation}
So we get the different directions of the branch curves at the
bifurcation point from the solutions of (\ref{4.121}) or
(\ref{4.122}). There are four possible cases:

Firstly, $A \neq 0,\,$ $\Delta =4(B ^2-AC )>0$, from Eq.
(\ref{4.121}) we
get two different solutions: $dv^1/dt\mid _{1,2}=(-B \pm \sqrt{B ^2-AC })/A $%
, which is shown in Fig. 2, where two topological defects meet and
then depart at the bifurcation point. Secondly, $A \neq 0,\,\Delta
=4(B ^2-A C )=0$, there is only one solution: $dv^1/dt=-B /A $,
which includes three important cases: (a) two topological defects
tangentially collide at the bifurcation point (Fig 3(a)); (b) two
topological defects merge into one topological defect at the
bifurcation point (Fig 3(b)); (c) one topological defect splits
into two topological defects at the bifurcation point (Fig 3(c)).
Thirdly, $A =0,\,C \neq 0,$ $\Delta =4(B ^2-A C )>0$, from Eq.
(\ref{4.122}) we have $dt/dv^1=0$ and $-2B /C $. There are two
important cases: (i) One topological defect splits into three
topological defects at the bifurcation point (Fig 4(a)); (ii)
Three topological defects merge into
one at the bifurcation point (Fig 4(b)). Finally, $A =C =0$, Eqs. (\ref{4.121}) and (%
\ref{4.122}) give respectively $dv^1/dt=0$ and $dt/dv^1=0$. This
case is obvious as in Fig. 5, which is similar to the third
situation.

In order to determine the branches directions of the remainder
variables, we will use the relations simply
\[
dv^A=f_1^Adv^1+f_t^Adt,\;\;\;\;\;A=2,3,...,n
\]
where the partial derivative coefficients $f_1^A$ and $f_t^A$ have given in (%
\ref{4.108}) and (\ref{4.109}). Then, respectively
\[
\frac{dv^A}{dv^1}=f_1^A+f_t^A\frac{dt}{dv^1}
\]
or
\begin{equation}  \label{4.126}
\frac{dv^A}{dt}=f_1^A\frac{dv^1}{dt}+f_t^A.
\end{equation}
where partial derivative coefficients $f_1^A$ and $f_t^A$ are
given by (\ref
{4.108}) and (\ref{4.109}). From this relations we find that the values of $%
dv^A/dt$ at the bifurcation point $(t^{*},z_i)$ are also possibly
different because (\ref{4.122}) may give different values of
$dv^1/dt$.

\subsection{The branch process at the higher degenerated point}

In the following, let us discuss the branch process at a higher
degenerated point. In the above subsection, we have analyzed the
case that the rank of the Jacobian matrix $[\partial \phi
/\partial v]$ of the equation (\ref {4.zero}) is $n-1$. In this
section, we consider the case that the rank of
the Jacobian matrix is $n-2$ (for the case that the rank of the matrix $%
[\partial \phi /\partial v]$ is lower than $n-2$, the discussion
is in the same way). Let the $(n-2)\times (n-2)$ submatrix
$J_2(\frac \phi v)$ of the Jacobian matrix $[\partial \phi
/\partial v]$ be
\[
J_2(\frac \phi v)=\left(
\begin{array}{cccc}
\phi _3^1 & \phi _4^1 & \cdots & \phi _n^1 \\ \phi _3^2 & \phi
_4^2 & \cdots & \phi _n^2 \\ \vdots & \vdots & \ddots & \vdots \\
\phi _3^{n-2} & \phi _4^{n-2} & \cdots & \phi _n^{n-2}
\end{array}
\right)
\]
and suppose that $\det J_2(\frac \phi v)|_{(t^{*},p_i)}\neq 0.$
With the same reasons of obtaining (\ref{4.107}), we can have the
function relations
\begin{equation}  \label{4.127}
v^A=f^A(v^1,v^2,t,\vec u ),\;\;\;\;\;A=3,4,...,n.
\end{equation}
For the partial derivatives $f_1^A$, $f_2^A$ and $f_t^A$, we can
easily derive the system similar to the equations (\ref{4.108})
and (\ref{4.109}), in which the three terms at the right hand of
can be figured out at the same
time. In order to determine the 2--order partial derivatives $f_{11}^A$, $%
f_{12}^A$, $f_{1t}^A$, $f_{22}^A$, $f_{2t}^A$ and $f_{tt}^A$, we
can use the method similar to the above mentioned. Substituting
the relations (\ref{4.127}) into the last two equations of the
system (\ref{4.phia}), we have the following two equations with
respect to
the arguments $v^1,\,\,v^2,\,\,t,\vec u $%
\begin{equation}  \label{4.bifa46}
\left\{
\begin{array}{l}
F_1(v^1,v^2,t, \vec u )=\phi ^{n-1}(v^1,v^2,f^3(v^1,v^2,t,\vec u
),\cdots ,f^n(v^1,v^2,t,\vec u ),t,\vec u )=0 \\
F_2(v^1,v^2,t,\vec u )=\phi ^n(v^1,v^2,f^3(v^1,v^2,t,\vec u
),\cdots ,f^n(v^1,v^2,t,\vec u ),t,\vec u )=0.
\end{array}
\right.
\end{equation}
Calculating the partial derivatives of the function $F_1$ and
$F_2$ with respect to $v^1$, $v^2$ and $t$, taking notice of
(\ref{4.127}) and using six similar expressions to (\ref{4.116})
and (\ref{4.117}), i.e.
\begin{equation}  \label{4.bifa48}
\frac{\partial F_j}{\partial v^1}\mid _{(t^{*},p_i)}=0,\ \quad \quad \frac{%
\partial F_j}{\partial v^2}\mid _{(t^{*},p_i)}=0,\ \quad \quad \frac{%
\partial F_j}{\partial t}\mid _{(t^{*},p_i)}=0,\ \quad \quad j=1,2,
\end{equation}
we have the following forms of Taylor expressions of $F_1$ and
$F_2$ in the neighborhood of $(t^{*},p_i)$
\[
F_j(v^1,v^2,t,\vec u )\approx
A_{j1}(v^1-p_i^1)^2+A_{j2}(v^1-p_i^1)(v^2-p_i^2)+A_{j3}(v^1-p_i^1)
\]
\[
\cdot
(t-t^{*})+A_{j4}(v^2-p_i^2)^2+A_{j5}(v^2-p_i^2)(t-t^{*})+A_{j6}(t-t^{*})^2=0
\]
\begin{equation}  \label{4.bifa49}
j=1,2.
\end{equation}
In the case of $A_{j1}\neq 0,A_{j4}\neq 0$, by dividing (\ref{4.bifa49}) by $%
(t-t^{*})^2$ and taking the limit $t\rightarrow t^{*}$, we obtain
two quadratic equations of $\frac{dv^1}{dt}$ and $\frac{dv^2}{dt}$
\begin{equation}  \label{4.bifa50}
A_{j1}(\frac{dv^1}{dt})^2+A_{j2}\frac{dv^1}{dt}\frac{dv^2}{dt}+A_{j3}\frac{%
dv^1}{dt}+A_{j4}(\frac{dv^2}{dt})^2+A_{j5}\frac{dv^2}{dt}+A_{j6}=0
\end{equation}
\[
j=1,2.
\]
Eliminating the variable $dv^1/dt$, we obtain a equation of
$dv^2/dt$ in the form of a determinant
\begin{equation}  \label{4.bifa51}
\left|
\begin{array}{cccc}
A_{11} & A_{12}Q+A_{23} & A_{14}Q^2+A_{15}Q+A_{16} & 0 \\ 0 &
A_{11} & A_{12}Q+A_{13} & A_{14}Q^2+A_{15}Q+A_{16} \\ A_{21} &
A_{22}Q+A_{23} & A_{24}Q^2+A_{25}Q+A_{26} & 0 \\ 0 & A_{21} &
A_{22}Q+A_{23} & A_{24}Q^2+A_{25}Q+A_{26}
\end{array}
\right| =0
\end{equation}
where $Q=dv^2/dt$, which is a $4th$ order equation of $dv^2/dt$
\begin{equation}  \label{4.bifa52}
a_0(\frac{dv^2}{dt})^4+a_1(\frac{dv^2}{dt})^3+a_2(\frac{dv^2}{dt})^2+a_3(%
\frac{dv^2}{dt})+a_4=0.
\end{equation}
Therefore we get different directions at the higher degenerated
point corresponding to different branch curves. The number of
different branch curves is four at
most. If the degree of degeneracy of the matrix $[\frac{\partial \phi }{%
\partial v}]$ is more higher, i.e. the rank of the matrix $[\frac{\partial
\phi }{\partial v}]$ is more lower than the present $(n-2)$ case,
the procedure of deduction will be more complicate. In general
supposing the rank of the matrix $[\frac{\partial \phi }{\partial
x}]$ be $(n-s)$, the number of the possible different directions
of the branch curves is $2^s$ at most.

At the end of this section, we conclude that there exist crucial
cases of branch processes in our theory of topological defect
system in TDGL model. This means that a topological defect, at the
bifurcation point, may split into several (for instance $m$)
topological defects along different branch curves with different
charges. Since the topological current is a conserved current, the
total quantum number of the splitting topological defects must
precisely equal to the topological charge of the original defect
i.e.
\[
\sum\limits_{j=1}^m\beta _{i_j}\eta _{i_j}=\beta _i\eta _i
\]
for fixed $i$. This can be looked upon as the topological reason
of the defect splitting. Here we should point out that such
splitting is a stochastic process, the sole restriction of this
process is just the conservation of the topological charge of the
topological defects during this splitting process. Of course, the
topological charge of each splitting defects is an integer.

\section*{Acknowledgment}

The author gratefully acknowledges the support of K. C. Wong
Education Foundation, Hong Kong. The author is also gratefully
indebted to Prof. Y. S. Duan for his warm-hearted helps and useful
discussion.

\section*{Figures' Captions}

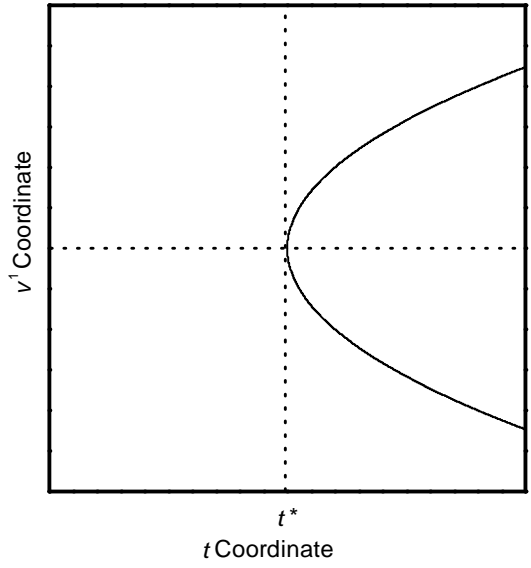
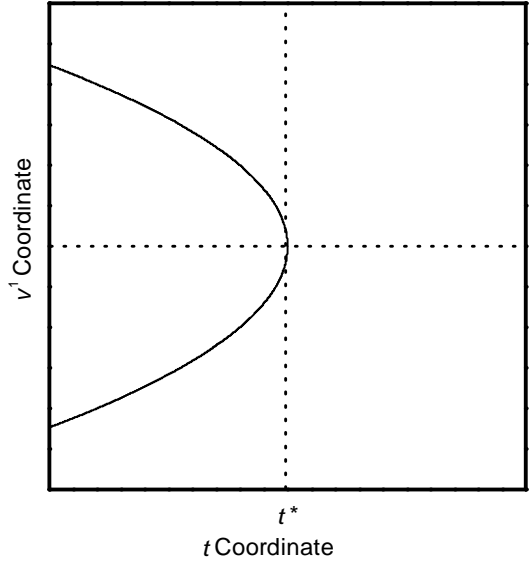
Fig. 1. (a) The creation of two topological defects. (b) Two
topological defects annihilate in collision at the limit point.

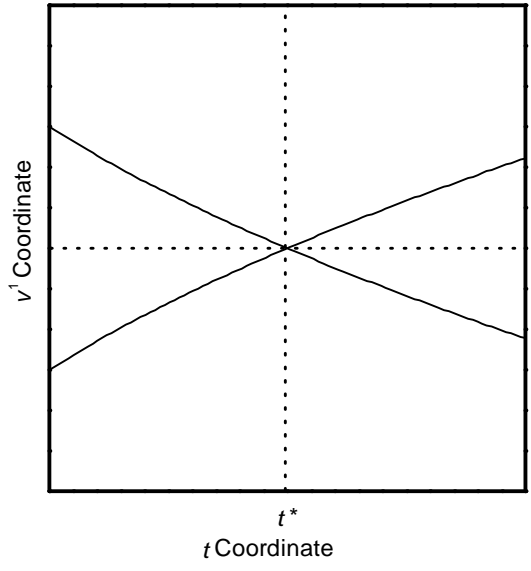
Fig. 2. Two topological defects collide with different directions
of motion at the bifurcation point.

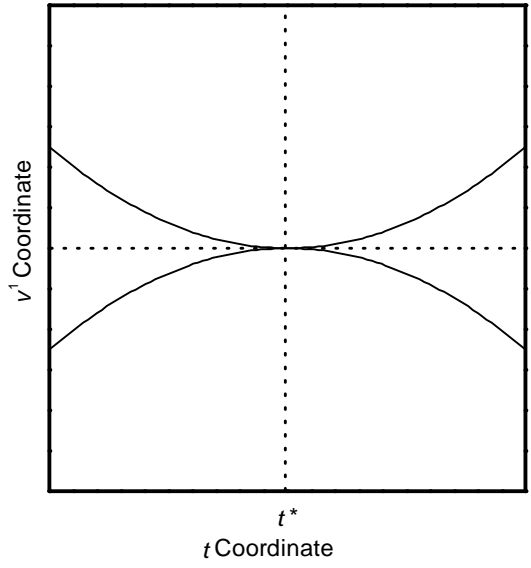
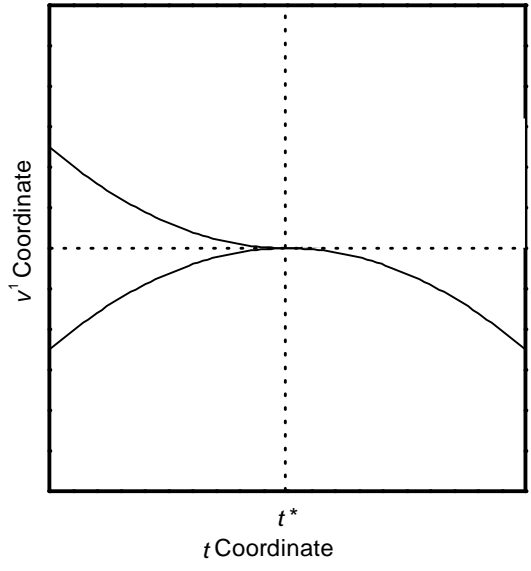
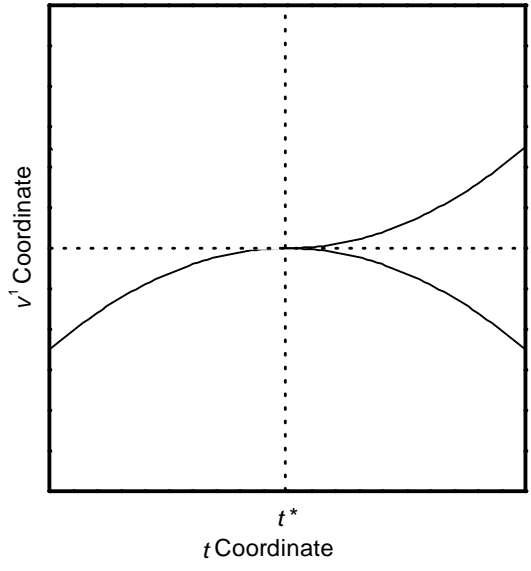
Fig. 3. Topological defects have the same direction of motion. (a)
Two topological defects tangentially collide at the bifurcation
point. (b) Two topological defects merge into one topological
defect at the bifurcation point. (c) One topological defect splits
into two topological defects at the bifurcation point.

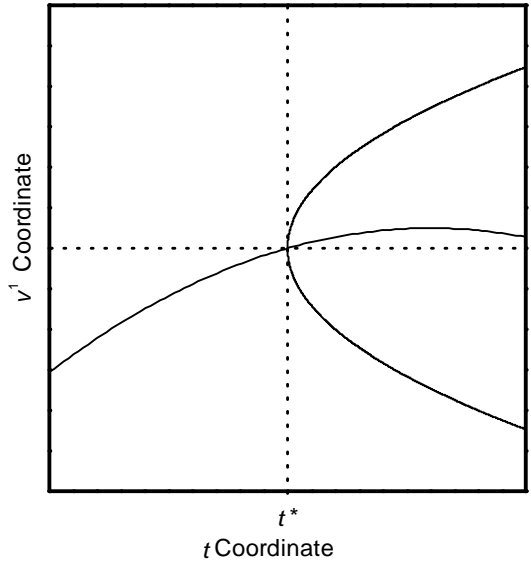
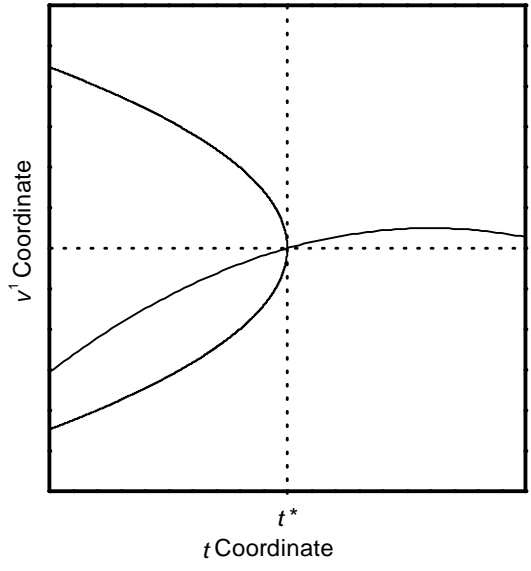
Fig. 4. (a) One topological defect splits into three topological
defects at the bifurcation point. (b) Three topological defects
merge into one topological defect at the bifurcation point.

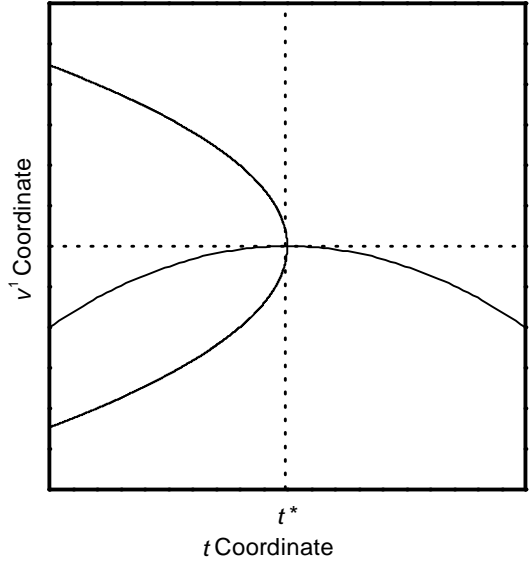
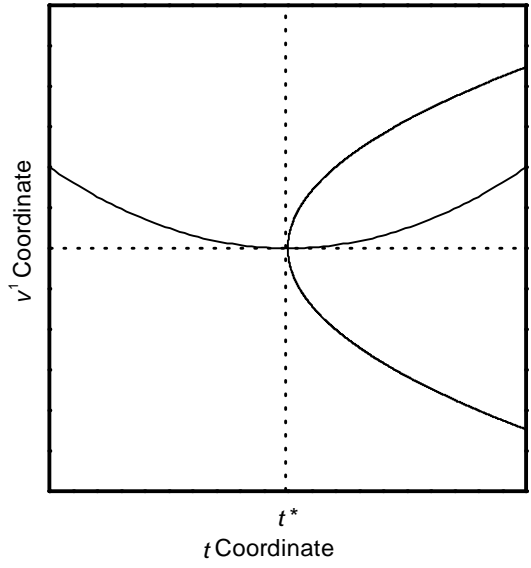
Fig. 5. This case is similar to Fig. 4. (a) Three topological
defects merge into one topological defect at the bifurcation
point. (b) One topological defect splits into three topological
defects at the bifurcation point.

\end{document}